\documentclass[prl,aps,twocolumn,tightenlines,floatfix,nofootinbib]{revtex4}

\usepackage{epsf}
\usepackage{graphicx}

\newcommand{\be}{\begin{equation}}
\newcommand{\ee}{\end{equation}}
\newcommand{\bear}{\begin{eqnarray}}
\newcommand{\eear}{\end{eqnarray}}
\newcommand{\n}{\hat{n}}

\newcommand{\rx}{{\rm x}}

\newcommand{\rn}{{\rm n}}
\newcommand{\rp}{{\rm p}}
\newcommand{\rnp}{{\rm np}}

\begin{document}

\title{On the stability of precessing superfluid neutron stars}

\author{K. Glampedakis$^1$, N. Andersson$^2$ \&  D.I. Jones$^2$}
\affiliation{$^1$SISSA/International School for Advanced Studies and INFN, via Beirut 2-4, 34014 Trieste, Italy \\
$^2$School of Mathematics, University of Southampton, Southampton SO17 1BJ, UK}

\begin{abstract}
We discuss a new superfluid instability occuring in the interior of mature neutron stars with implications
for freely precessing neutron stars. This short-wavelength instability is  similar to
the instability which is responsible for the formation of 
turbulence in superfluid Helium. Its existence raises serious questions about
our understanding of neutron
star precession and complicates 
attempts to constrain neutron star interiors using such observations.
\end{abstract}

\maketitle


{\em Introduction}.--- Neutron stars tend to be extremely stable rotators, with stability that
sometimes rivals that of the best atomic clocks. Yet a
growing sample of pulsars exhibit spin irregularities, like glitches and timing noise.
They may also be undergoing free precession. From the theory point of view, one might expect
precession to be generic. Nevertheless, for reasons
still to be understood,  compelling evidence for long-period precession has only been found in the
timing data of a few pulsars. The best candidate is PSR B1828-11 \cite{1828}
which exhibits a $\sim 500~\mbox{d} $ high-quality periodicity, with an amplitude of a few degrees.
The paucity of precessing neutron stars is one of the mysteries of pulsar physics.
To explain why precession is so rare is difficult. After all, a description of even a modestly
realistic neutron star requires the fusion of much of modern theoretical physics.
One would need to account for strong gravity, supranuclear density matter, superfluidity/superconductivity
and potentially very strong magnetic fields. 

In the standard picture of a mature neutron star the bulk of the neutrons are superfluid and
rotate by forming a dense array of vortices. Meanwhile the outer core protons
are expected to form a type~II superconductor, with the magnetic flux
carried by fluxtubes. The coupling between these two distinct fluid
components is usually assumed to have the same form as in the 
case of superfluid Helium, see \cite{sidery06} for a recent discussion. However, this model is based on the assumption that the
neutron vortex array is (locally) straight. This may not be the case. In a body that undergoes
a more complex motion one might expect to find that the vortices get tangled up to form a state of
superfluid turbulence. In Helium, the formation of a vortex tangle is assumed to
follow the onset of an instability in the vortex array \cite{glab}. It has recently been suggested that
an analogue of this so-called Donnelly-Glaberson instability may be relevant for neutron stars 
\cite{turbulent,sidery07,peralta1} . If this is the case, one would expect it to have interesting repercussions
for neutron star precession. In this Letter
we confirm this expectation by demonstrating that short-wavelength instabilities
are generic in precessing superfluid neutron stars.


{\em Plane wave analysis}. ---
Our main objective is to investigate whether analogues of the Donnelly-Glaberson instability
are likely to occur in a neutron star interior. Our analysis is based on the standard
two-fluid picture, where the superfluid neutrons are dynamically distinguished from
a conglomerate of comoving superconducting protons and normal electrons. We will loosely
refer to the latter as the ``protons'' in the following. Variables associated with each fluid will be
labelled by $\rx =\{\rn,\rp\} $.
The smooth-averaged hydrodynamics of the system is governed by two coupled Euler-type equations, see
\cite{sidery06} for more details.
In a frame rotating  with angular velocity $\Omega^i$ we have 
\bear
&& D_t^\rn v_i^\rn  + \nabla_i \psi_\rn =  2 \epsilon_{ijk} v^j_\rn \Omega^k
 + f_i^{\rm mf}
\label{eulern}
\\
&& D_t^\rp v_i^\rp + \nabla_i \psi_\rp =  2 \epsilon_{ijk} v^j_\rp \Omega^k
- f_i^{\rm mf}/x_\rp
 + \nu_{\rm ee} \nabla^2 v_i^\rp
\label{eulerp}
\eear
Here
the fluid velocities are denoted by $v^i_\rx $, we have introduced the convective derivatives
$D_t^\rx = \partial_ t  + v^j_\rx \nabla_j$
and $x_\rp = \rho_\rp/\rho_\rn $ is the density fraction.
The scalars  $\psi_\rx$ are the sums of specific
chemical potentials and the gravitational potential \cite{sidery06}. For simplicity,
we assume that both fluids are incompressible, i.e. we have
$\nabla_i v^i_\rx = 0 $. In the interest of clarity, we also ignore the entrainment effect in this
study.
A key property of the system is that neutrons and protons are coupled via mutual friction, a force
$f_i^{\rm mf}$ mediating the interaction between the quantized
neutron vortices and the proton fluid/magnetic fluxtubes.
The standard expression for this force is, see \cite{sidery06}, 
\be
f_i^{\rm mf} =  {\cal B} \epsilon_{ijk} \epsilon^{kml} \hat{\omega}^j_\rn \omega_m^\rn w^\rnp_l
+  {\cal B}^\prime \epsilon_{ijk} \omega^j_\rn w^k_\rnp
\label{mf}
\ee
where $w^i_\rnp = v^i_\rn -v^i_\rp$ and
the neutron vorticity is given by $\omega^i_\rn = \epsilon^{ijk} \nabla_j v_k^\rn $.
 A ``hat'' denotes a unit vector. This form for the mutual friction force results from balancing the
 Magnus force that acts on the neutron vortices and a resistive ``drag'' force which represents the
 interaction between the vortices and the charged fluid \cite{sidery06}. Representing the drag force
 by a dimensionless coefficient $\cal R$, one finds that
\be
{\cal B} = \frac{{\cal R}}{1 + {\cal R}^2}, \qquad
\mbox{and} \qquad  {\cal B}^\prime = \frac{{\cal R}^2}{1 + {\cal R}^2}
\ee

In the most commonly considered case, the mutual friction arises from scattering of electrons
off the vortex's intrinsic magnetic field. This leads to a relatively weak coupling, with
\cite{als,sidery06},
\be
{\cal R} \approx 4\times 10^{-4}~\Rightarrow~ {\cal B}^\prime \approx {\cal B}^2 \quad
{\cal B} \approx {\cal R} \ll 1
\label{weak}
\ee
It is, however, not  established that it  is this limit that applies. Hence,
one must also consider the case of strong coupling which follows from taking $ {\cal R} \gg 1 $.
This translates into
\be
{\cal B} \ll 1, \qquad {\cal B}^\prime \approx 1 -{\cal B}^2
\ee
The strong coupling limit is relevant if the interaction between neutron vortices and
fluxtubes is efficient \cite{ruderman,link03, link06}, if there is a fluxtube cluster associated with each neutron vortex
\cite{sed2}, or if there is significant vortex pinning \cite{shaham}
(in the limit ${\cal R} \to \infty$ the
vortices can be considered as perfectly ``pinned'').

Returning to the Euler equations (\ref{eulern})--(\ref{eulerp}), only the proton equation
contains a  shear viscosity term. This is because the dominant process is expected to
be electron-electron scattering. The upshot of this is that the neutron fluid is not
directly affected by shear viscosity.
The relevant viscosity coefficient $\nu_\mathrm{ee}$ has been estimated in \cite{viscous}.
For a uniform density star with  $M=1.4M_\odot$,  $R=10$~km
and $x_\rp =0.1$ (the canonical values we will use later) we have
$\nu_\mathrm{ee} \approx 10^7 (T/10^8 \mathrm{K})^{-2}$~cm$^2$/s.

We consider perturbations of eqns.~(\ref{eulern}) and (\ref{eulerp}) for a background configuration
where both fluids rotate rigidly with $v^i_{\rx 0} = \epsilon^{ijk} ( \Omega^\rx_j -\Omega_{j}) x_k $.
By allowing for an arbitrary orientation of the  angular velocity vectors, this
configuration can represent the standard free precession modes of a two-fluid star \cite{swc,dij}.
We then linearise the Euler equations, focussing
on short-wavelength motion by making the standard plane-wave decomposition
\be
\delta v^i_\rx = A^i_\rx \, e^{ i (\sigma t + k_j x^j)}, \quad A^i_\rx = \mbox{constant}
\label{plane}
\ee
and similarly for all other variables.
Since we expect the flow along the background vortex array to play a central role \cite{turbulent,sidery07}, we
carry out the perturbation calculation in the neutron frame. That is, we take
$\Omega^i= \Omega^i_\rn = \Omega_\rn \n^i$.
In order to simplify the problem, without any real loss of generality \cite{sidery07}, we only consider waves
propagating along the vortices, i.e. $k^i = k_\parallel \n^i$. Then the fact that we have assumed the
fluids to be incompressible means that the waves are transverse,
$\n_i A^i_\rx =0 $. 
 After some algebra, cf. \cite{sidery07} for a similar analysis, the perturbed
versions of (\ref{eulern}) and (\ref{eulerp}) lead to a $4\times4$ system, the determinant of which
provides the dispersion relation for short-wavelength waves. 
A detailed analysis of the problem will be provided elsewhere. Here we focus 
on the modes that may become unstable.

Let us first consider the weak drag limit. Then we find
a mode with frequency (with viscous corrections of order $1/k_\parallel$)
\be
\sigma \approx  2\Omega_\rn + ( i {\cal B} - {\cal B}') \left  (\, 2\Omega_\rn - k_\parallel w_\parallel
\, \right )
\label{mode_w}\ee
Here $w_\parallel$ represents the relative linear flow along the (background) neutron vortex array.
 In our case we have
$w_\parallel = - \n^i \epsilon_{ijk} \Omega^j_\rp x^k$. In the local analysis we have taken
 $w_\parallel$ to be constant. Hence it is 
clear that our analysis is only consistent for short wavelength motion.
Anyway, from (\ref{mode_w}) we see that the
system is unstable ($\mbox{Im } \sigma <0 $) if
\be
w_\parallel> 2\Omega_\rn/k_\parallel 
\label{crit_w}\ee
As discussed in \cite{sidery07}, the solution (\ref{mode_w}) represents inertial waves in the neutron fluid.
This instability is the exact analogue of the Donnelly-Glaberson instability in Helium \cite{glab}, and hence
its existence should come as no real surprise. As in Helium, one would expect the onset of
the instability to lead to the formation of tangled vortices, reconnection and superfluid turbulence.
Since turbulence alters the form of the macroscopic
mutual friction force \cite{turbulent,peralta1}, it is not yet clear how the system will evolve
once the unstable waves grow to large amplitude.

As far as we are aware, the strong drag problem has not been considered previously.
Interestingly, there are unstable modes also in this case. The nature of the instability is, however, 
more complex. In the strong drag limit, with $\mathcal{B}=0$ and $\mathcal{B}'=1$, we find a 
mode with frequency
\bear
&& \sigma \approx \Omega_\rn \left ( 1 -\frac{1}{x_\rp} \right ) + k_\parallel w_\parallel + i { \nu_\mathrm{ee} k_\parallel^2 \over 2}
- \left\{  { \Omega^2_\rn (1+ x_\rp)^2 \over x_\rp } \right. \nonumber \\
&&  \left. - { 2\Omega_\rn k_\parallel w_\parallel \over x_{\rp}} - {\nu_\mathrm{ee}^2 k_\parallel^4  \over 4}
- i {(1-x_\rp) \over x_\rp}  \nu_\mathrm{ee} k_\parallel^2 \Omega_\rn  \right\}^{1/2}
\label{mode_s}
\eear
This result clearly shows that there will be unstable waves
(representing coupled inertial waves in the neutron/proton fluids). In the inviscid
($\nu_\mathrm{ee}=0$) limit the instability is active  provided that
\be
w_\parallel > \frac{\Omega_\rn (1+ x_\rp)^2}{2 k_\parallel x_{\rm p}}
\label{crit_s}
\ee
As in the weak drag case, one would expect the onset of this instability to lead to tangled vortices.


{\em Implications for precessing neutron stars}. ---
In order to discuss the implications of the above results we need to make contact
between our background configuration and the global precession motion.
Fortunately, this is straightforward. The precession
of a two-component neutron star model, including mutual friction coupling, has already
been discussed in \cite{swc}. The simplest
model consists of two components that rotate rigidly. The neutron component is assumed spherical
with moment of inertia $I_\rn$. At the same time, the protons (including the crust)
are assumed to be slightly deformed in such a way that $I_\rp = I_1=I_2=I_3/(1+\epsilon)$ with $\epsilon \ll 1$
(in a principal coordinate system where the deformation axis is along $\hat{x}_3$).
When perturbed away from alignment of the
two rotation axes, $\Omega_\rx^i$, the crust precesses with a certain frequency and observable wobble angle
$\theta_w$ (the angle between the deformation axis, $\hat{x}_3$, and the total angular momentum axis) \cite{dij}.

The plane-wave analysis is consistent for the precessing system provided that the two
rotation vectors $\Omega_\rx^i $ can be considered fixed. This is true as long as
the precession period $P_{\rm pr}$ is significantly longer than the
timescale associated with the local waves. As already mentioned, the wavelength of the waves we consider must also
be short enough that $w_\parallel$ can be treated as a constant.
If these conditions hold then we are simply considering local perturbations of a given precession model.
In order the check whether this system is locally stable we only need to work out $w_\parallel$ from the
precession solution. If an instability is present, then the precession solution must be considered
questionable. It certainly cannot be the case that the two components rotate rigidly, a key assumption
in the standard analysis \cite{swc}.

{\em Weak drag slow precession}. ---
In the weak drag limit, there exists a long period precession solution
that is slowly damped by mutual friction. For this solution we have \cite{swc}
\be
P_{\rm pr} \approx P/\epsilon \ , \quad \mbox{and} \quad
t_d \approx { P \over 2\pi {\cal B}   \epsilon} { I_\rp \over I_\rn}
\ee
where $P_\mathrm{pr}$ is the precession period, $t_d$ is the
damping time are $P$ is the rotation period of the star. We then find that
\be
w_\parallel \approx  2 \pi \epsilon \theta_w  x_2/P
\label{inst_w}\ee
where $x_2(<R)$ is one of the coordinates associated with crust system.
This estimate can be used in (\ref{crit_w}) to show that all waves
with wavelength ($\lambda = 2\pi/k_\parallel$) shorter than
\be
\lambda_\mathrm{max}  =
5\times 10^{-4} \left( {\theta_w \over 1^\circ } \right)\left( {\epsilon \over 10^{-8} } \right)
\left( { R \over 10^6 \mathrm{cm} } \right) \ \mathrm{cm}
\ee
are unstable. However, there must
be a short wavelength cut-off for the instability.
To make progress it would seem natural to assume that our analysis
becomes invalid once the wavelength is so short that the fluid description is no longer relevant.
Then it seems reasonable to use something like
\be
\lambda_\mathrm{min} \approx 100 d_\rn \approx \left( {P \over 1 \mathrm{s} } \right)^{1/2} \ \mathrm{cm}
\ee
where $d_\rn$ is the intervortex spacing.
Since we need to have $\lambda_\mathrm{max}>\lambda_\mathrm{min}$ in order to argue that the instability is
relevant we see that we must have
\be
\left( { \theta_w \over 1^\circ} \right) > 1900 \left( {P \over 1 \mathrm{s} } \right)^{1/2}
\left( {\epsilon \over 10^{-8} } \right)^{-1} \left( { R \over 10^6 \mathrm{cm} } \right)^{-1}
\label{thetalimit1}\ee
What does this result tell us? It suggests that, if the drag is weak, the superfluid instability is unlikely to play a role in
slowly spinning
systems. For the archetypal precessor PSR B1828-11 \cite{1828} the spin period is
0.4~s and in order to have precession with the observed period one would need $\epsilon \approx 10^{-8}$.
It is then clear from (\ref{thetalimit1}) that precession with a wobble angle of a few degrees is safely in the
stable regime. Nevertheless, the weak drag result is not without interest. Consider for example a millisecond
pulsar with a maximally strained crust. From (\ref{thetalimit1}) we see that if the spin period is 1~ms, then
precession with $\theta_w$ larger than a degree would be unstable provided that $\epsilon > 6\times 10^{-7} $.
Since the theoretically predicted range for crustal deformations has  $\epsilon <  10^{-5}$ \cite{haskell}, we see that
our result puts a constraint on slow precession in very fast spinning neutron stars.

{\em Strong drag fast precession}. ---
In the strong drag limit, the relevant precession solution is such that  \cite{swc}
\be
P_{\rm pr} \approx {I_\rp \over I_\rn} P \ , \quad \mbox{and} \quad
t_d \approx { P \over 2\pi {\cal B} } { I_\rp \over I_\rn}
\label{fastprec}\ee
and we find that
\be
w_\parallel \approx  {2 \pi \theta_w  x_2 \over P} {I_\rn \over I_\rp}
\ee
The (inviscid) instability criterion (\ref{crit_s}) then
implies that waves with wavelength shorter than
\be
 \lambda_\mathrm{max} = 2\times 10^5 \left( {\theta_w \over 1^\circ } \right)
 \left( { R \over 10^6 \mathrm{cm} } \right) \ \mathrm{cm}
\label{lambda_2}\ee
(we have assumed  $I_\rp/I_\rn \approx x_\rp$) will be unstable. 

\begin{figure}
\centerline{\includegraphics[height= 6.5cm,clip]{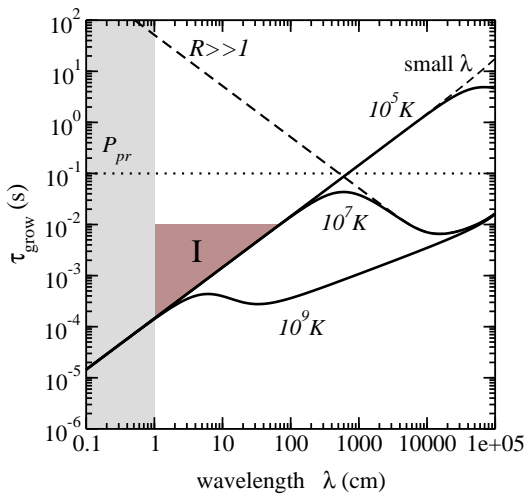}}
\caption{Growth timescales $\tau_{\rm grow}(\lambda) $ for unstable superfluid waves in a
precessing neutron star. This particular model has ${\cal R}=10^3$, representing the strong drag regime,
$P=1$~s and $\theta_w=1^\circ$.  The dotted horizontal line
shows the fast precession period $P_\mathrm{pr}$ that follows if we take $x_\rp = I_\rp/I_\rn=0.1$.
We compare our numerical results for three different core temperatures (solid lines) to two approximations. 
The short wavelength approximation and the $\mathcal{R}\gg1$ approximation (\ref{mode_s}) are 
both indicated by dashed lines.
The region where a short-wavelength 
instability operates (I) for $T=10^7$~K is highlighted.
Finally, the $\lambda < \lambda_{\rm min} $ region where we assume that the hydrodynamical description fails is
shaded.}
\label{fig1}
\end{figure}

However, as is clear from (\ref{mode_s}), the unstable strong drag 
modes are affected by viscosity. 
To unveil the detailed behaviour we
have solved the dispersion relation numerically for a range of parameter values. Typical results are 
shown in Figure~\ref{fig1}. This figure shows $\tau_\mathrm{grow}$ as a function of $\lambda$ and illustrates
how the importance of shear viscosity 
varies with temperature. The results for  
core temperatures $T=10^9$~K and $T=10^7$~K show a clear transition from a regime where the inviscid approximation
to (\ref{mode_s}) is valid (above $\lambda \approx 20$~cm and $10^4$~cm, respectively) to a short wavelength 
regime where viscosity alters the result. However, a surprising feature appears as one proceeds towards shorter wavelengths 
for a fixed temperature. As $k_\parallel$ becomes large, it turns out that
there is a cancellation of the leading order viscosity terms, cf. (\ref{mode_s}). For short wavelengths, the
mode frequencies are instead accurately (with errors 
of order $1/k_\parallel$) described by 
(\ref{mode_w}) with $\mathcal{B}'=1$. Hence, for $\lambda \ll \lambda_\mathrm{max}$ 
the short wavelength modes grow on a timescale given by 
\be
\tau_\mathrm{grow} \approx {\lambda \over 2 \pi \mathcal{B} | w_\parallel | }  
\ee 
For typical parameters, we have 
\be
\tau_\mathrm{grow} > 140  \left( { x_\rp \over 0.1} \right)
\left( {\theta_w \over 1^\circ } \right)^{-1} \left( { P \over 1\ \mathrm{s} } \right) 
\left( { \mathcal{R} \over 10^3 } \right) \left( { \lambda \over R } \right) \ \mathrm{s}
\label{tau_est}\ee
For consistency the unstable waves need to grow on a timescale that is short 
compared to the precession period.
If we require (say) $\tau_\mathrm{grow} < 0.1 P_\mathrm{pr}$, then we have 
\be
\lambda < 70 \left( {\mathcal{R} \over 10^3} \right)^{-1} \left( {\theta_w \over 1^\circ } \right) \ \mathrm{cm}
\ee
The corresponding instability region is indicated by an I in Fig.~\ref{fig1}. 
Moreover, in order to have $ \lambda >  \lambda_\mathrm{min}$ (noting that the short wavelength cut-off remains as
in the weak drag case) we must have
\be
\mathcal{R} < 7 \times10^4 \left( {\theta_w \over 1^\circ } \right) \left( { P \over 1\ \mathrm{s} } \right)^{-1/2}
\ee
This shows that the short-wavelength instability 
constrains a wide range of fast precession models.
From the results in Fig.~\ref{fig1} it is also clear that there may exist a medium wavelength instability regime
(well approximated by (\ref{mode_s})). This is relevant for temperatures above $10^7$~K, and could well lead to the 
fastest growing instability in young neutron stars. 

{\em Brief discussion}. ---
In this Letter we have demonstrated that short wavelength superfluid instabilities may operate in freely precessing neutron stars.
In the weak drag regime, the instability affects only rapidly spinning stars that have significantly
deformed crusts. PSR B1828-11, the currently best candidate precessor, lies well within the stable 
regime. In contrast, our results have serious implications for systems in the 
strong drag regime. We predict that these systems will suffer local instabilities, possibly 
leading the formation of superfluid turbulence, for a wide range of the relevant parameter space. 
This calls into question the standard precession model, which is based on two co-existing fluids rotating 
as solid bodies \cite{swc}, and any conclusions drawn from it.
 In particular, one would note Link's argument
\cite{link03,link06} that the
coupling between vortices and fluxtubes ought to lead to fast precession according to (\ref{fastprec}).
Since this is contradicted
by the observed slow precession of PSR B1828-11, Link suggests that our understanding of the neutron star core physics
is wrong and that the protons would actually form a type~I superconductor (without fluxtubes).
Our results add an element of doubt. We have essentially shown that the strong drag 
fast precession solution may be
inconsistent for a
neutron star spinning at the rate of  PSR B1828-11. If the precessing motion  triggers
a range of unstable short wavelength waves then the original solid-body assumption that led to
(\ref{fastprec}), cf. \cite{swc}, cannot hold. The precession problem may  thus be more complex
than usually assumed, and a consistent description of fast precession must
properly include superfluid wave dynamics
and potential turbulence.

\acknowledgements
This work was supported by 
PPARC/STFC via grant number PP/E001025/1.



\begin{thebibliography}{10}

\bibitem{1828}
I.H. Stairs, A.G. Lyne and S.L. Shemar, Nature (London) {\bf 406}, 484 (2000)

\bibitem{sidery06}
N. Andersson, T. Sidery and G.L. Comer, MNRAS {\bf 368}, 162 (2006)

\bibitem{glab}
W.I. Glaberson, W.W. Johnson and R.M. Ostermeier, Phys. Rev. Lett., {\bf 33} 1197 (1974)

\bibitem{turbulent}
N. Andersson, T. Sidery and G.L. Comer, to appear in MNRAS (preprint astro-ph/0703257)

\bibitem{sidery07}
T. Sidery, N. Andersson and G.L. Comer, submitted to MNRAS (preprint astro-ph/0706.0672)

\bibitem{peralta1} 
C. Peralta, A. Melatos, M. Giacobello and A. Ooi, Ap. J. {\bf 635} 1224 (2005); ibid. {\bf 651} 1079 (2006)

\bibitem{als}
M.A. Alpar, S.A. Langer and J.A. Sauls, Ap. J. {\bf 282} 533 (1984)

\bibitem{ruderman}
M. Ruderman, T. Zhu and K. Chen, Ap. J. {\bf 492}  267 (1998) 

\bibitem{link03}
B. Link, Phys. Rev. Lett., {\bf 91},  101101 (2003)

\bibitem{link06}
B. Link, Astron. Astrophys.  {\bf 458}, 881 (2006)

\bibitem{sed2}
A.D. Sedrakian and D.M. Sedrakian, Ap. J. {\bf 447} 305 (1995)

\bibitem{shaham}
J. Shaham, Ap. J., {\bf 214}, 251 (1977)

\bibitem{viscous}
N. Andersson, G.L. Comer and K. Glampedakis, Nucl. Phys. A {\bf 763}, 212 (2005)

\bibitem{swc}
A. Sedrakian, I. Wasserman and J.M. Cordes, Ap. J. {\bf 524}, 341 (1999)

\bibitem{dij}
D.I. Jones and N. Andersson, MNRAS {\bf 324} 811 (2001)

\bibitem{haskell} 
B. Haskell, D.I. Jones and N. Andersson, MNRAS {\bf 373} 1423 (2006)
\end{thebibliography}
\end{document}